\documentclass[a4paper,12pt]{article}
\usepackage{amssymb}
\usepackage{latexsym}

\usepackage{epsfig}
\usepackage{graphics}

\usepackage{amsmath,amsthm,amsfonts,amssymb}

\newcommand{\be}{\begin{eqnarray}}
\newcommand{\ee}{\end{eqnarray}}
\newcommand{\bea}{\begin{eqnarray}}
\newcommand{\eea}{\end{eqnarray}}
\newcommand{\nn}{\nonumber}

\newcommand{\nk}{\noindent}

\begin{document}

\begin{titlepage}
\begin{flushright}
hep-th/0108013\\ UA/NPPS-09-2001
\end{flushright}
\begin{centering}
\vspace{.8in}
{\large {\bf General Brane Dynamics with $^{(4)}R$ term in the
Bulk}}
\\

\vspace{.5in} {\bf Georgios Kofinas\footnote{gkofin@phys.uoa.gr}}
\\

\vspace{0.3in}
University of Athens\\ Physics Department\\
Nuclear and Particle Physics Section\\
Panepistimioupolis, Ilisia GR 157 71\\ Athens, Greece\\
\end{centering}

\vspace{0.7in}
\begin{abstract}
\nk A general analysis of the induced brane dynamics is performed
when the intrinsic curvature term is included in the action. Such
a term is known to cause dramatic changes and is generically
induced by quantum corrections coming from the bulk gravity and
its coupling with matter living on the brane. The induced brane
dynamics is shown to be the usual Einstein dynamics coupled to a
well defined modified energy-momentum tensor. In cosmology,
conventional general relativity revives for an initial era whose
duration depends on the value of the five-dimensional Planck mass.
Violations of energy conditions may be possible, as well as matter
inhomogeneities on the brane in $(A)dS_{5}$ or Minkowski
backgrounds. A new anisotropic cosmological solution is given in
the above context. This solution, for a fine-tuned
five-dimensional cosmological constant, exhibits an intermediate
accelerating phase which is followed by an era corresponding to a
4D perfect fluid solution with no future horizons.
\\
\\
\\
\\
\\
\\
\\
\\
July, 2001
\end{abstract}
\end{titlepage}

\newpage

\baselineskip=18pt
\section*{1 \,\,\,Introduction}
\hspace{0.8cm} Brane cosmologies consist cosmological realizations
of string theories in which some underlying features are often
minimized. Ordinary matter fields living on the brane are
represented by open-string excitations, while gravitons by closed
string loops, which not being trapped on the brane, enable the
gravitons to penetrate into the higher dimensional bulk. Thus,
gravity is fundamentally a higher dimensional effect and the bulk
Einstein equations are generally used to deduce the dynamics on
the brane. For a given bulk metric, the induced metric on a
specific brane is certainly uniquely defined. Since there is
basically no preferred position for a brane in a given bulk, the
knowledge of the particular induced metric is irrelevant. However,
the knowledge of the induced brane gravitational dynamics,
irrespectively of the brane position, becomes a fact of particular
importance. When the brane has codimension one and the effective
low-energy theory in the bulk is higher dimensional gravity, such
a reduction is possible \cite{maeda}. A more fundamental
description of the physics that produces the brane could include
\cite{sundrum} higher order terms in a derivative expansion of the
effective action, such as a term for the scalar curvature of the
brane, and higher powers of curvature tensors on the brane. A
brane action that contains powers of the brane curvature tensors
has also been used in the context of the $AdS/CFT$ correspondence
(e.g. \cite{maldacena}) to regularize the action of a bulk $AdS$
space which diverges when the radius of the $AdS$ space becomes
infinite. If the dynamics is governed not only by the ordinary
five-dimensional Einstein-Hilbert action, but also by the
four-dimensional Ricci scalar term induced on the brane, new
phenomena appear. In \cite{dvali1, dvali2} it was observed that
the localized matter fields on the brane (which couple to bulk
gravitons) can generate via quantum loops a localized
four-dimensional worldvolume kinetic term for gravitons (see also
\cite{capper, adler, zee, khuri}). That is to say,
four-dimensional gravity is induced from the bulk gravity to the
brane worldvolume by the matter fields confined to the brane. It
was also shown that an observer on the brane will see correct
Newtonian gravity at distances shorter than a certain crossover
scale, despite the fact that gravity propagates in extra space
which was assumed there to be flat with infinite extent. At larger
distances, the force becomes higher-dimensional. The first
realization of the induced gravity scenario in string theory was
presented in \cite{kiritsis}. Furthermore, new closed string
couplings on Dp-branes for the bosonic string were found in
\cite{corley}. These couplings are quadratic in derivatives and
therefore take the form of induced kinetic terms on the brane. For
the graviton in particular these are the induced Einstein-Hilbert
term as well as terms quadratic in the second fundamental tensor.
The inclusion of an intrinsic curvature term is naturally expected
to affect the cosmological expansion of the Universe. In
\cite{collins, shtanov, nojiri, deffayet, myung} the
Friedmann-like equations for particular cosmological examples were
discussed in this context. In \cite{deffayet} two isotropic
cosmological solutions were found (in Minkowski bulk) close to the
usual FRW cosmology for small enough Hubble radius, while far
beyond the cross-over scale, the Universe enters a fully 5D regime
or to a self-inflationary expansion arising without including any
effective cosmological constant on the brane. With
five-dimensional Planck mass $M_{5}$ being of the order of $TeV$
\cite{dimopoulos} the above scenario seems phenomenologically not
very viable at least from a cosmological and astronomical
perspective \cite{dvali1}. However, in \cite{dvali3} a $TeV$
string scale together with one extra compact dimension of
astronomical size can circumvent all phenomenological
difficulties. Further, in \cite{dvali4} a framework with an
infinite-volume flat 5D space was proposed in which the quantum
gravity scale can be as low as $10^{-3}\,eV$ without conflicting
with any of the existing laboratory, astrophysical or cosmological
bounds. An observer on the brane can see conventional gravity from
a lower distance up to astronomically large distances; outside
this range gravity becomes five-dimensional. If $M_{5} \sim
10\,MeV$, then \cite{defdvali, deffayet} a self-accelerated 4D
universe predicts modification of gravitational laws at scales
comparable with the present cosmological horizon and can be in
agreement with the recent Supernovae and acoustic peaks of Cosmic
Microwave Background radiation. Earlier discussions on lowering
the string scale relative to $M_{4}$ were given in \cite{witten,
lykken}. In \cite{kiritsis} a relevant analysis with $^{(4)}R$
term introduced a new threshold scale, due to the width of a
non-zero thickness brane; at distances larger than this scale,
physics is four-dimensional, while at shorter distances irregular
deviations appear. Systems similar to the one studied in the
present paper have been also studied in \cite{nunez} where some
discussion of the role played by the induced curvature term within
the context of the $AdS/CFT$ correspondence has been given. The
situation with an Einstein curvature term added to the brane
action, but without imposing the reflectional symmetry between the
two sides of the brane, has been discussed in \cite{vilenkin}.
\par
In the present work we are treating the general relativistic case
with no restrictions on the metric form, when the above intrinsic
curvature term is included. The additional term is geometrical in
nature and thus, creates modified geometrical terms beyond the
Einstein ones in the induced (brane) equations. This is the line
of work followed in the relevant cosmological models
\cite{collins, shtanov, nojiri, deffayet} and makes the further
investigation of the induced dynamics difficult. We show that
these induced equations can assume a usual Einstein equation form
with additional matter content. This enables us to treat the brane
dynamics with the methods of conventional general relativity. More
specifically, the energy-momentum tensor in these equations is
split into the common brane energy-momentum tensor plus additional
terms which are all multiplied by one of the characteristic scales
of the theory, $M_{5}^{3}/M_{4}^{2}$. In the cosmological context,
conventional general relativity revives for an initial era which
depends on the value of the above scale. An ordinary matter on the
brane produces terms in these effective equations which may cause
violations of the energy conditions. Such violations arising from
a different context also appeared in \cite{kehagias}. We have
performed major part of the analysis in $n+1$ spacetime
dimensions. In an $(A)dS$ or Minkowski bulk the conservation of
energy on the brane does not necessarily imply the existence of
only spatially homogeneous universes (as this happens when
$^{(4)}R$ is not included \cite{maeda}), i.e. inhomogeneous
perfect fluid brane solutions might exist in the above exact
bulks. In section 3, we incorporate the above formulation to the
class of four-dimensional spatially homogeneous spacetimes and
find, as an example, a new anisotropic cosmological
four-dimensional solution. The boundary condition used for the
embedding of the brane in the bulk is through the vanishing of the
electric part of the five-dimensional Weyl tensor. For a
fine-tuned five-dimensional cosmological constant, this solution
possesses an accelerating phase after the conventional
four-dimensional evolution, followed by another 4D perfect fluid
solution with no event horizons. Since this contains the parameter
$M_{5}$, one could try to compare this with the well-known value
of the quadrupole moment of the CMB radiation and thus set
restrictions on the value of the string scale. Finally, in section
4, we conclude and speculate on possible generalizations.

\section*{2 \,\,General analysis with intrinsic curvature Ricci scalar}
\hspace{0.8cm} We start with a $(n+1)$-dimensional theory and a
$(n-1)$-brane $\Sigma$ embedded in $(n+1)$-dimensional spacetime
$M$. Capital Latin letters $A,B,...=0,1,...,n$ will denote full
spacetime, lower Greek $\mu,\nu,...=0,1,...,n-1$ run over brane
worldvolume, while lower Latin ones span some $(n-1)$-dimensional
spacelike surfaces foliating the brane, i.e. $i,j,...=1,...,n-1$.
For convenience, we can quite generally, choose a coordinate $y$
such that the hypersurface $y=0$ coincides with the brane. Our
primary interest lies in $n=4$ where the brane is supposed to
represent our universe. The total action for the system is taken
to be: \be
S&=&\frac{1}{2\kappa_{n+1}^{2}}\int_{M}\sqrt{\varepsilon_{n+1}\,^{(n+1)}g}\,\,(^{(n+1)}R-2\Lambda_{n+1})d^{n+1}x+
\frac{1}{2\kappa_{n}^{2}}\int_{\Sigma}\sqrt{\varepsilon_{n}\,^{(n)}g}\,\,(^{(n)}R-2\Lambda_{n})d^{n}x\nn\\&&+
\int_{M}\sqrt{\varepsilon_{n+1}\,^{(n+1)}g}\,\,L_{n+1}^{mat}\,d^{n+1}x+
\int_{\Sigma}\sqrt{\varepsilon_{n}\,^{(n)}g}\,\,L_{n}^{mat}\,d^{n}x.
\label{action} \ee For clarity, we have separated the cosmological
constants $\Lambda_{n+1}$, $\Lambda_{n}$ from the rest matter
contents $L_{n+1}^{mat}$, $L_{n}^{mat}$ of the bulk and the brane
respectively. $\Lambda_{n}/\kappa_{n}^{2}$ can be interpreted as
the brane tension of the standard Dirac-Nambu-Goto action, or as
the sum of a brane worldvolume cosmological constant and a brane
tension. For possible treatment of different than usual signatures
of the metrics we have allowed the signs $\varepsilon_{n+1}$,
$\varepsilon_{n}$. If the bulk is $(A)dS$, the $(n+1)$-dimensional
Weyl tensor vanishes and also $^{(n+1)}T_{AB}=0$. If the bulk is
Minkowski, additionally $\Lambda_{n+1}=0$.
\par
From the dimensionful constants $\kappa_{n+1}^{2}$,
$\kappa_{n}^{2}$ the Planck masses $M_{n+1}$, $M_{n}$ are defined
as:
 \be
  \kappa_{n+1}^{2}=8\pi
G_{(n+1)}=M_{n+1}^{-(n-1)}\,\,\,\,\,,\,\,\,\,\,
\kappa_{n}^{2}=8\pi G_{(n)}=M_{n}^{-(n-2)}, \label{planck} \ee
with $M_{n+1}$, $M_{n}$ having dimensions of (length)$^{-1}$.
Then, a distance scale $r_{c}$ is defined as :
 \be
r_{c}\equiv\frac{\kappa_{n+1}^{2}}{\kappa_{n}^{2}}=\frac{M_{n}^{n-2}}
{M_{n+1}^{n-1}}\,.
 \label{distancescale}
 \ee
Varying (1) with respect to the bulk metric $g_{AB}$, we obtain
the equations
\be
^{(n+1)}G_{AB}=-\Lambda_{n+1}g_{AB}+\kappa_{n+1}^{2}\,(^{(n+1)}T_{AB}+\,^{(loc)}T_{AB}\,\delta(y))\,,
\label{varying}
 \ee
 where
\be
^{(loc)}T_{AB}\equiv-\frac{1}{\kappa_{n}^{2}}\,\sqrt{\frac{\varepsilon_{n}\,^{(n)}g}
{\varepsilon_{n+1}\,^{(n+1)}g}}\,\,(^{(n)}G_{AB}-\kappa_{n}^{2}\,^{(n)}T_{AB}+
\Lambda_{n}h_{AB})
 \label{tlocal}
\ee is the localized energy-momentum tensor of the brane.
$^{(n+1)}G_{AB}$, $^{(n)}G_{AB}$ denote the Einstein tensors
constructed from the bulk and the brane metrics respectively.
Clearly, $^{(n)}G_{AB}$ acts as an additional source term for the
brane through $^{(loc)}T_{AB}$. The tensor $h_{AB}=g_{AB}-\epsilon
_{y}n_{A}n_{B}$ is the induced metric on the hypersurfaces $y=$
constant, with $n^{A}$ the normal vector on these.
\par
The way the $y$-coordinate has been defined allows us to write at
least in the neighborhood of the brane the $(n+1)$-line element
as: \be
ds_{(n+1)}^{2}=g_{AB}dx^{A}dx^{B}=\varepsilon_{t}N^{2}dt^{2}+g_{ij}dx^{i}dx^{j}+\varepsilon_{y}\,b^{2}dy^{2}\,,
\label{lineelement}
\ee
where $N, g_{ij}, b$ are generally
functions of $t,x^{i},y$. The splitting of the brane metric into
space and time parts allows the choice of zero shift in
(\ref{lineelement}).
\par
According to the metric (\ref{lineelement}), the Einstein
equations (\ref{varying}) of the bulk space are split into the
following set of equations of the canonical analysis \cite{wald}:
\be
K_{\mu;\nu}^{\nu}-K_{;\mu}=\varepsilon_{y}\kappa_{n+1}^{2}\,b\,\,^{(n+1)}T_{\mu}^{y}
\label{linear} \ee \be
K_{\nu}^{\mu}K_{\mu}^{\nu}-K^{2}+\varepsilon_{y}\,^{(n)}R=2\varepsilon_{y}(\Lambda_{n+1}-\kappa_{n+1}^{2}\,^{(n+1)}T_{y}^{y})
\label{quadratic} \ee \be
K_{\nu}^{\mu}\,\acute{}+bKK_{\nu}^{\mu}-\varepsilon_{y}b\,\,^{(n)}R_{\nu}^{\mu}&+&
\varepsilon_{y}g^{\mu\lambda}b_{;\lambda\nu}=
-\varepsilon_{y}\kappa_{n+1}^{2}b\,\Big(\,^{(loc)}T_{\nu}^{\mu}-\frac{^{(loc)}T}{n-1}\,\delta_{\nu}^{\mu}\,\Big)\delta(y)-\nn\\&&
-\varepsilon_{y}\kappa_{n+1}^{2}b\,^{(n+1)}T_{\nu}^{\mu}
+\frac{\varepsilon_{y}b}{n-1}(\kappa_{n+1}^{2}\,^{(n+1)}T
-2\Lambda_{n+1})\delta_{\nu}^{\mu}, \label{dynamical} \ee where
$K_{AB}=h_{A}^{C}h_{B}^{D}\nabla_{C}n_{D}$ is the extrinsic
curvature of the hypersurfaces $y$=constant. In the above
equations, $K,\,^{(n+1)}T,\,^{(loc)}T$ denote the traces of
$K_{B}^{A},\,^{(n+1)}T_{B}^{A},\,^{(loc)}T_{B}^{A}$ (indices
raised by $g^{AB}$), the semicolon stands for covariant
differentiation with respect to the induced metric $g_{\mu\nu}$,
while throughout, prime and dot mean partial derivatives with
respect to $y$ and $t$ respectively.
\par
Isolating the singular part of equations (\ref{dynamical}) we
obtain the modified (due to $^{(n)}G_{\nu}^{\mu}$)
Israel-Darmois-Lanczos-Sen conditions \cite{israel, darmois,
lanczos, sen}
 \be
[K_{\nu}^{\mu}]=-\varepsilon_{y}\kappa_{n+1}^{2}b_{o}\,\left(^{(loc)}T_{\nu}^{\mu}-
\frac{^{(loc)}T}{n-1}\delta_{\nu}^{\mu}\right)\,, \label{israel}
\ee where the bracket means discontinuity of the quantity across
$y=0$, and $b_{o}=b(y=0)$. Hereafter, we consider a
$\mathbf{Z}_{2}$ symmetry on reflection around the brane and thus
(\ref{israel}) becomes \be
^{(n)}G_{\nu}^{\mu}=\kappa_{n}^{2}\,^{(n)}T_{\nu}^{\mu}-\Lambda_{n}\delta_{\nu}^{\mu}+
\varepsilon_{y}\alpha\,(\overline{K}\,_{\nu}^{\mu}-\overline{K}\delta_{\nu}^{\mu})\,,
 \label{eqn1}
 \ee
where
$\overline{K}\,_{\nu}^{\mu}=K_{\nu}^{\mu}(y=0^{+})=-K_{\nu}^{\mu}(y=0^{-})$
and $\alpha \equiv 2sgn(b_{o})/r_{c}$.
\par
These equations resemble Einstein equations on the brane but
unfortunately, they contain the undetermined geometrical quantities
$\overline{K}\,_{\nu}^{\mu}$. Fortunately however, we can do
better based on a geometrical identity, namely Gauss
equation, i.e.
\be
^{(n)}R^{A}_{BCD}=\,^{(n+1)}R^{M}_{NKL}h_{M}^{A}h_{B}^{N}h_{C}^{K}h_{D}^{L}
+\varepsilon_{y}\,(K_{C}^{A}K_{BD}-K_{D}^{A}K_{BC}).
 \label{gauss}
\ee From the above relation, taking suitable contractions to
construct the $n$, $(n+1)$-dimensional Einstein tensors, and
making use of the bulk Einstein equations, we get \be
^{(n)}G_{AB}&=&\varepsilon_{y}\,(KK_{AB}-K_{AC}K_{B}^{C})+
\frac{\varepsilon_{y}}{2}(K_{D}^{C}K_{C}^{D}-K^{2})h_{AB}-
\frac{n-2}{2}\Lambda_{n+1}h_{AB}+\nn\\&&+\,\frac{n-2}{n-1}\,\kappa_{n+1}^{2}\,\Big(\,^{(n+1)}T_{CD}h_{A}^{C}h_{B}^{D}
+\Big(\varepsilon_{y}\,^{(n+1)}T_{CD}n^{C}n^{D}-\frac{1}{n}\,^{(n+1)}T_{C}^{C}\Big)h_{AB}\Big)-\nn\\&&
-\,\varepsilon_{y}C_{ACBD}n^{C}n^{D}\,, \label{contraction} \ee
where $C^{A}_{BCD}$ is the Weyl tensor of $^{(n+1)}R^{A}_{BCD}$.
The parallel to our brane, components of the preceding equations
give \be
^{(n)}G_{\nu}^{\mu}&=&\varepsilon_{y}\,(\overline{K}\,\overline{K}\,_{\nu}^{\mu}-\overline{K}\,_{\lambda}^{\mu}\overline{K}\,_{\nu}^{\lambda})
+\frac{\varepsilon_{y}}{2}\,(\overline{K}\,_{\lambda}^{\kappa}\overline{K}\,_{\kappa}^{\lambda}-\overline{K}\,^{2})\delta_{\nu}^{\mu}
-\frac{n-2}{n}\,\Lambda_{n+1}\delta_{\nu}^{\mu}+\nn\\&&
+\,\frac{n-2}{n-1}\,\kappa_{n+1}^{2}\,\Big(\,^{(n+1)}\overline{T}\,_{\nu}^{\mu}
+\Big(\,^{(n+1)}\overline{T}\,_{y}^{y}-\frac{^{(n+1)}\overline{T}}{n}\Big)\,\delta_{\nu}^{\mu}\Big)
-g^{\kappa \mu}\,\overline{C}\,^{y}_{\kappa y \nu }\,\,,
\label{eqn2} \ee where we have put a bar over $^{(n+1)}T_{B}^{A}$,
$C^{y}_{\kappa y \nu}$ to show explicitly that all the quantities
in (\ref{eqn2}) are evaluated at $y=0$. Equations
(\ref{contraction}), (\ref{eqn2}) do not contain any singular part
since the distributional part of the system has been extracted as
the boundary condition for the bulk imposed by equation
(\ref{eqn1}).
\par
Equations (\ref{eqn2}) are independent from (\ref{eqn1}), so we
can get additional information on $\overline{K}\,_{\nu}^{\mu}$ by
equating their right hand sides: \be
\overline{K}\,_{\lambda}^{\mu}\,\overline{K}\,_{\nu}^{\lambda}-\overline{K}\,\overline{K}\,_{\nu}^{\mu}
+\frac{1}{2}\,(\overline{K}\,^{2}-\overline{K}\,_{\lambda}^{\kappa}\,\overline{K}\,_{\kappa}^{\lambda})\,\delta_{\nu}^{\mu}
+\alpha(\overline{K}\,_{\nu}^{\mu}-\overline{K}\delta_{\nu}^{\mu})=\mathcal{T}_{\nu}^{\mu}\,,
\label{alg1} \ee where \be
\mathcal{T}_{\nu}^{\mu}&=&\varepsilon_{y}\,\Big(\Lambda_{n}-\frac{n-2}{n}\,\Lambda_{n+1}\Big)\delta_{\nu}^{\mu}
-\varepsilon_{y}\,\kappa_{n}^{2}\,^{(n)}T_{\nu}^{\mu}+\nn\\&&
+\varepsilon_{y}\,\frac{n-2}{n-1}\,\kappa_{n+1}^{2}\,\Big(\,^{(n+1)}\overline{T}\,_{\nu}^{\mu}
+\Big(\,^{(n+1)}\overline{T}\,_{y}^{y}-\frac{^{(n+1)}\overline{T}}{n}\Big)\,\delta_{\nu}^{\mu}\Big)
-\varepsilon_{y}\,\overline{\textsf{E}}^{\,\mu}_{\,\nu}\,,
 \label{energy}
 \ee
 and $\textsf{E}_{AB}=C_{ACBD}n^{A}n^{B}$ is the electric part of
 the Weyl tensor.
\par
Equation (\ref{alg1}) is an algebraic equation for
$\overline{K}\,_{\nu}^{\mu}$ in terms of
$\mathcal{T}_{\nu}^{\mu}$, which we will try to solve. Taking the
trace of (\ref{alg1}) and plugging back in the same equation, we
obtain equivalently \be
\overline{K}\,_{\lambda}^{\mu}\,\overline{K}\,_{\nu}^{\lambda}-(\overline{K}-\alpha)\overline{K}\,_{\nu}^{\mu}
+\frac{\alpha}{n-2}\overline{K}\,\delta_{\nu}^{\mu}=\mathcal{T}_{\nu}^{\mu}-
\frac{\mathcal{T}_{\lambda}^{\lambda}}{n-2}\,\delta_{\nu}^{\mu}\,.
\label{alg2} \ee Now, setting
 \be
L_{\nu}^{\mu}\equiv
\overline{K}\,_{\nu}^{\mu}-\frac{\overline{K}-\alpha}{2}\,\delta_{\nu}^{\mu}\,,
\label{l} \ee equation (\ref{alg2}) gets the form \be
L_{\lambda}^{\mu}L_{\nu}^{\lambda}-\frac{L^{2}}{(n-2)^{2}}\,\delta_{\nu}^{\mu}
=\mathcal{T}_{\nu}^{\mu}-\frac{(n-1)\alpha^{2}+(n-2)\mathcal{T}_{\lambda}^{\lambda}}{(n-2)^{2}}\,\delta_{\nu}^{\mu}\,,
\label{lll} \ee where $L\equiv L_{\mu}^{\mu}$. Supposed that
(\ref{lll}) has been solved, substituting
$\overline{K}\,_{\nu}^{\mu}$ back in (\ref{eqn1}) in terms of the
matter contents $\mathcal{T}\,_{\nu}^{\mu}$, we will have strictly
a system of Einstein equations for the brane metric with
additional matter terms besides the conventional ones, i.e. \be
^{(n)}G_{\nu}^{\mu}=\kappa_{n}^{2}\,^{(n)}T_{\nu}^{\mu}-\Big(\Lambda_{n}
+\varepsilon_{y}\,\frac{n-1}{n-2}\alpha^{2}\Big)\,\delta_{\nu}^{\mu}+
\varepsilon_{y}\alpha\Big(L_{\nu}^{\mu}+\frac{L}{n-2}\,\delta_{\nu}^{\mu}\Big)\,.
\label{einstein} \ee An effective cosmological constant arises
which exists even for $\Lambda_{n}=0$. The various terms appeared
in the right hand side of equation (\ref{einstein}) consist the
effective energy-momentum tensor, which presumably can violate
some energy conditions, even if the conventional matter terms do
not. When $M_{4}$ in the action (\ref{action}) becomes much larger
than $M_{5}$, then $\alpha\rightarrow 0$ and the above equation
reduces to the expected four-dimensional General Relativity (at
least whenever the quantity multiplying $\alpha$ does not
diverge).
\par
Due to the block-diagonal form of the $n$-part of metric
(\ref{lineelement}), we have that
$\overline{K}\,_{i}^{0}=\overline{K}\,_{0}^{i}=0$ and thus
$L_{i}^{0}=L_{0}^{i}=0$. Then, from (\ref{lll}) it arises that
$\mathcal{T}_{i}^{0}=\mathcal{T}_{0}^{i}=0$. Furthermore, the
system (\ref{lll}) can be decomposed into the following set of
equations:
 \be
L_{l}^{i}L_{j}^{l}=((L_{0}^{0})^{2}-\mathcal{T}_{0}^{0})\,\delta_{j}^{i}+\mathcal{T}_{j}^{i}\,,
\label{dec1}
\ee
\be
L_{i}^{i}=-L_{0}^{0}\pm
\sqrt{(n-2)^{2}(L_{0}^{0})^{2}-(n-2)(n-3)\mathcal{T}_{0}^{0}+(n-2)\mathcal{T}_{i}^{i}+(n-1)\alpha^{2}}\,\,.
\label{dec2}
 \ee
  The unknown matrices $L_{j}^{i}$ in (\ref{dec1})
are $(n-1)$-dimensional and if solved in terms of $L_{0}^{0}$ and
$\mathcal{T}_{\nu}^{\mu}$, then (\ref{dec2}) will set an algebraic
equation for $L_{0}^{0}$. In order for a solution to exist, it has
to be
$((L_{0}^{0})^{2}-\mathcal{T}_{0}^{0})\delta_{j}^{i}+\mathcal{T}_{j}^{i}$
and the quantity under the square root of (\ref{dec2})
non-negative.
\par
To assure the existence of solutions for equation
(\ref{dec1}), we assume the situation where $\mathcal{T}_{j}^{i}$
has $n-1$ (real) eigenvalues $\tau_{1},...,\tau_{n-1}$, not
necessarily distinct and its minimal polynomial has simple roots
only. (Of course, the case with $\mathcal{T}_{j}^{i}$ having $n-1$
distinct eigenvalues is also included). Then, there exists some
invertible ($x^{\mu}$-dependent) matrix $P_{j}^{i}$ which
diagonalizes $L_{j}^{i}$, i.e.
\be
L_{j}^{i}=(P^{-1})_{k}^{i}\,\widetilde{L}_{l}^{k}\,P_{j}^{l}\,,
\label{diagonalize1}
\ee
 where the components in this new frame
are \be
\widetilde{L}_{j}^{i}=diag\Big(\pm\sqrt{(L_{0}^{0})^{2}+\tau_{1}-\mathcal{T}_{0}^{0}}\,\,,...,
\pm\sqrt{(L_{0}^{0})^{2}+\tau_{n-1}-\mathcal{T}_{0}^{0}}\,\Big)\,.
\label{components}
 \ee
 The various $\pm$ appeared are independent
each other. Then, equation (\ref{dec2}) gives
 \be
&\pm&\sum_{i=1}^{n-1}\,\sqrt{(L_{0}^{0})^{2} +
\tau_{i}-\mathcal{T}_{0}^{0}}\,\pm\,L_{0}^{0}=\nn\\&&=
\sqrt{(n-2)^{2}(L_{0}^{0})^{2}-(n-2)(n-3)\mathcal{T}_{0}^{0}+
(n-2)\sum_{i=1}^{n-1}\,\tau_{i}+(n-1)\alpha^{2}} \label{implicit}
\ee with all the $\pm$'s being independent. The above equation is
not easily solved for $L_{0}^{0}$ in terms of the matter, but if
this is done, then (\ref{einstein}) become well-defined Einstein
equations for the brane with modified energy-momentum tensor.
\par
Of particular importance is the subcase with
\underline{$\mathcal{T}_{j}^{i}$ isotropic}, i.e.
$\mathcal{T}_{j}^{i}=\tau\,\delta_{j}^{i}$ (as it is seen from
(\ref{energy}) this happens, for example, when the brane and/or
the bulk contain untilted perfect fluids and
$\overline{\textsf{E}}^{\,i}_{\,j}$ is proportional to the
identity. Then, the solution of (\ref{dec1}) given by
(\ref{diagonalize1}) and (\ref{components}) is \be
L_{j}^{i}=S\,(P^{-1})_{k}^{i}\,E_{l}^{k}\,P_{j}^{l}\,,
\label{diagonalize2} \ee where
 \be
S=\sqrt{(L_{0}^{0})^{2}+\tau-\mathcal{T}_{0}^{0}}\,\,,
\label{s1}
\ee
 \be
 E_{j}^{i}=diag(+1,...,+1,-1,...,-1)\,,
 \label{e}
\ee
with $n_{+}\geq0$ signs $+1$, $n_{-}\geq0$ signs $-1$, and
$n_{+}+n_{-}=n-1$. We have to discern two cases.
\par
 Let \underline{$n_{+},n_{-}\neq0,1$}. Then, equation (\ref{dec2})
 (or equivalently (\ref{implicit})) is a quadratic for $(L_{0}^{0})^{2}$
 and supplies the following explicit solution for $L_{0}^{0}$\,:
\be L_{0}^{0}&=&\oplus
\frac{1}{\sqrt{8n_{+}n_{-}(n_{+}-1)(n_{-}-1)}}\,[-2(n_{+}-1)
(n_{-}-1)(n-1-4n_{+}n_{-})
\mathcal{T}_{0}^{0}+\nn\\&&+2n_{+}n_{-}(3n-5-4n_{+}n_{-}) \tau
+(n-1)(n-1-2n_{+}n_{-}) \alpha^{2} \pm
|n_{+}-n_{-}|\,A]^{\frac{1}{2}}
 \label{solution1}
 \ee
 where
 \be
A=&&[\,-4(n_{+}-1)(n_{-}-1)\mathcal{T}_{0}^{0}\,\,(\,(n-2)\mathcal{T}_{0}^{0}+(n-1)\alpha^{2}\,)
+\nn\\&&+\,4n_{+}n_{-}\tau
\,\,(\,(n-2)\tau+(n-1)\alpha^{2}\,)\,+\,(n-1)^{2}\alpha^{4}\,\,]^{\frac{1}{2}}\,.
\label{aa} \ee The quantity $S$ in (\ref{diagonalize2}) becomes
\be S=
\frac{1}{\sqrt{8n_{+}n_{-}(n_{+}-1)(n_{-}-1)}}&[&-2(n_{+}-1)
(n_{-}-1)(n-1)\mathcal{T}_{0}^{0}-2n_{+}n_{-}(n-3)\tau+
\nn\\&&+(n-1)(n-1-2n_{+}n_{-}) \alpha^{2} \pm
|n_{+}-n_{-}|\,A\,\,]^{\frac{1}{2}}.
 \label{sb}
  \ee
It is assumed that the matter contents are such that the square
roots appeared in (\ref{solution1}), (\ref{aa}), (\ref{sb}) and
the right hand side of (\ref{implicit}) are all well-defined. The
sign $\oplus$ in (\ref{solution1}) means that an overall $+$ or
$-$ sign can be taken independently of the other $\pm$ of
(\ref{solution1}), (\ref{sb}) which however go together.
\par
Let \underline{some of $n_{+},n_{-}$ take the value $0$ or $1$}.
(Obviously, the case of our primary interest $n=4$ is included
here). Then, (\ref{dec2}) (or (\ref{implicit})) has vanishing
coefficient of $(L_{0}^{0})^{4}$ and gives : \be
L_{0}^{0}=\pm\frac{1}{2}\,[(3n-5-4n_{+}n_{-})\mathcal{T}_{0}^{0}-(n-1-4n_{+}n_{-})\tau+(n-1)\alpha^{2}]\,/B\,,
\label{solution2} \ee where \be
B=[-2(n_{+}-1)(n_{-}-1)(n-1)\mathcal{T}_{0}^{0}&&+2\,n_{+}n_{-}(1-n_{+}n_{-})\tau+\nn\\&&+\,(n-1)(n-1-2n_{+}n_{-})\alpha^{2}\,]^{\frac{1}{2}}\,,
\label{b}
 \ee
 and from (\ref{s1}):
 \be
S=\frac{1}{2}\,|(n-3)\mathcal{T}_{0}^{0}+(n-1)(\tau+\alpha^{2})|\,/B\,.
\label{sbb}
\ee
In this case, we are remained with the square root of (\ref{b}) and
that of the right hand side of (\ref{implicit}) to be
well-defined.
\newline
We observe that since $n_{+}, n_{-}$ insert
equations (\ref{solution1}), (\ref{aa}), (\ref{sb}),
(\ref{solution2}), (\ref{b}), (\ref{sbb}) in a symmetric way, the interchange of the number of $+1$'s with that of
$-1$'s does not affect the equations.
\par
Gathering together the above results for the ``isotropic'' case,
we rewrite equations (\ref{einstein}) for the previously found
$L_{0}^{0}$, $S$ (equations (\ref{solution1}), (\ref{sb}),
(\ref{solution2}), (\ref{sbb})) as: \be
^{(n)}G_{0}^{0}=\kappa_{n}^{2}\,^{(n)}T_{0}^{0}-\Big(\Lambda_{n}+
\varepsilon_{y}\,\frac{n-1}{n-2}\,\alpha^{2}\Big)+\varepsilon_{y}\alpha\,\frac{n-1}{n-2}\,\Big(L_{0}^{0}+\frac{n_{+}-n_{-}}{n-1}\,S\Big)\,,
\label{einstein1} \ee \be
^{(n)}G_{j}^{i}=\kappa_{n}^{2}\,^{(n)}T_{j}^{i}-\Big(\Lambda_{n}+\varepsilon_{y}\,
\frac{n-1}{n-2}\,\alpha^{2}\Big)\delta_{j}^{i}+\varepsilon_{y}\alpha
S E_{j}^{i}+\frac{\varepsilon_{y}\alpha}{n-2}\,
(L_{0}^{0}+(n_{+}-n_{-})S)\,\delta_{j}^{i}\,, \label{einstein2}
 \ee
where now the indices $i,j$ refer to the new frame. Thus, the
inclusion of the term $^{(n)}R$, instead of creating additional
difficulties, has brought a convenient decomposition of the matter
terms. First, standard energy-momentum tensor enters without
having made (as in \cite{binetruy, maeda}) any choice for the
brane tension $\Lambda_{4}$ in terms of $M_{4},\,M_{5}$ (namely
$\Lambda_{4}=3\alpha^{2}/2$). Note that if $^{(4)}R$ is not
included in the action, then for $\Lambda_{4}=0$, ordinary
energy-momentum terms cannot arise. Furthermore, in that case,
$\Lambda_{4}$ has to be positive in order for $\kappa_{4}^{2}$ to
be positive. Second, the additional matter terms (which rather
appear here as square roots instead of squares of the
four-dimensional energy-momentum tensor) are \textit{all}
multiplied by the energy scale $\alpha$. We will come back on this
later on, but this already sounds promising in reviving standard
general relativistic cosmologies, by simply constraining the value
of Planck mass $M_{5}$. Without including the intrinsic curvature
term, conventional four-dimensional Einstein gravity arises in the
low energy world, i.e. whenever the characteristic energy scale of
the matter, $\kappa_{4}^{2}\,^{(4)}T_{\nu}^{\mu}$, is much lower
than $\Lambda_{4}$, when additionally to the above referred
condition for $\Lambda_{4}$, the constants
$\Lambda_{5},\,\Lambda_{4}$ are fine-tuned \cite{binetruy, maeda}.
However, in \cite{maeda} it was stated that if simply
$\kappa_{5}\rightarrow 0$, keeping $\kappa_{5}^{4}\Lambda_{4}$
finite, common four-dimensional gravity arises. This is not
exactly correct, since under the above conditions, the brane
constant $\Lambda_{4}$ (even for a Minkowski bulk) does not enter
the induced Einstein equations as it is, but multiplied by $1/2$.
\par
If we concentrate on the case $n=4$, there are only two
different situations: (i) $n_{+}=3$,
$n_{-}=0$, and (ii) $n_{+}=2$, $n_{-}=1$.
\newline
(i) \,\, From (\ref{solution2}), (\ref{sbb}) we have
\be
L_{0}^{0}=\pm\frac{1}{2\sqrt{3}}\,\frac{7\mathcal{T}_{0}^{0}-3\tau+3\alpha^{2}}
{\sqrt{4\mathcal{T}_{0}^{0}+3\alpha^{2}}}\,\,\,\,\,\,\,\,,\,\,\,\,\,\,\,\,
S=\frac{1}{2\sqrt{3}}\,\frac{|\mathcal{T}_{0}^{0}+3\tau+3\alpha^{2}|}{\sqrt{4\mathcal{T}_{0}^{0}+3\alpha^{2}}}\,.
\label{4l1}
\ee
\newline
(ii) \,\, Similarly, we obtain
 \be
L_{0}^{0}=\pm\frac{1}{2}\,\frac{-\mathcal{T}_{0}^{0}+5\tau+3\alpha^{2}}
{\sqrt{-4\tau-3\alpha^{2}}}\,\,\,\,\,\,\,\,,\,\,\,\,\,\,\,\,
S=\frac{1}{2}\,\frac{|\mathcal{T}_{0}^{0}+3\tau+3\alpha^{2}|}{\sqrt{-4\tau-3\alpha^{2}}}\,.
\label{4l2}
\ee
The only conditions which have to be satisfied
here are those concerning the square roots of the denominators in
(\ref{4l1}), (\ref{4l2}).
\par
Taking the jump across $y=0$ of the linear equations
(\ref{linear}) and using (\ref{israel}), we obtain the usual
conservation equations for the brane \be ^{(n)}T_{\nu \,;\,
\mu}^{\mu}=0\,, \label{conservation} \ee just like the case where
the $^{(4)}R$ term was not present (assuming that there is no
energy flow from the brane towards the bulk and vice-versa, i.e.
$^{(n+1)}T_{\mu}^{y}=0$). From equations (\ref{einstein}) written
in the new frame, (\ref{conservation}), and the contracted Bianchi
identities $^{(n)}G_{\nu\,;\,\mu}^{\mu}=0$, it arises that \be
L_{\nu\,;\,\mu}^{\mu}+\frac{L_{\,;\,\nu}}{n-2}=0\,. \label{ll} \ee
Whenever $L_{\nu}^{\mu}$ has been found in terms of the matter,
the above equations (\ref{ll}) will impose restrictions on the
form of the matter admitted.
\par
The quantity $\overline{\textsf{E}}^{\,\mu}_{\,\nu}$ in
(\ref{energy}) carries the influence of non-local gravitational
degrees of freedom in the bulk onto the brane and makes the brane
equations (\ref{einstein}) not to be, in general, closed. This
means that there are bulk degrees of freedom which cannot be
predicted from data available on the brane. One has to solve the
field equations in the bulk in order to determine
$\textsf{E}^{\,\mu}_{\,\nu}$ on the brane. However, the symmetry
properties of $\textsf{E}_{AB}$ imply that in general, this can be
decomposed irreducibly \cite{roy} with respect to a chosen
4-velocity field $u^{\mu}$, in terms of a non-local energy density
on the brane, a non-local anisotropic stress, and a non-local
energy flux on the brane. One way for making (\ref{einstein})
closed is to set $\overline{\textsf{E}}^{\,\mu}_{\,\nu}=0$ as a
boundary condition of the propagation equations in the bulk space.
Then, equations (\ref{ll}), assuming additionally
$^{(5)}T_{AB}=0$, are written equivalently as: \be
3\dot{L_{0}^{0}}+(n_{+}-n_{-})\dot{S}+\frac{\dot{\gamma}}{\gamma}L_{0}^{0}-
Sg^{jl}\dot{g}_{li}E_{j}^{i}=0\,, \label{ls1} \ee \be
L_{0\,,\,i}^{0}+(n_{+}-n_{-})S_{,i}+2S_{,j}E_{i}^{j}+2\frac{N_{,j}}{N}
\,(SE_{i}^{j}-L_{0}^{0}\delta_{i}^{j})+2S\,(\,^{(3)}\Gamma_{lj}^{j}E_{i}^{l}-\,^{(3)}\Gamma_{ij}^{l}E_{l}^{j}\,)=0\,,
\label{ls2} \ee where $L_{0}^{0}\,,S$ are given by (\ref{4l1}) or
(\ref{4l2}).
\par
If the matter content of the brane is an untilted perfect fluid,
i.e. $^{(n)}T_{\mu \nu}=(\rho +p)u_{\mu}u_{\nu}+pg_{\mu\nu}$,
where $u^{0}=\frac{1}{N}$ (all other components of $u^{A}$ are
zero) and $p=p(\rho)$ is the equation of state, then
(\ref{conservation}) is known for metric (\ref{lineelement}) to be
written equivalently as: \be
\dot{\rho}+(\rho+p)\frac{\dot{\gamma}}{2\gamma}=0\,, \label{ro}
\ee \be p_{,i}+(\rho+p)\frac{N_{,i}}{N}=0\,, \label{pe} \ee where
$\gamma \equiv det(g_{ij})$. Substituting
$\frac{\dot{\gamma}}{\gamma}$ and $\frac{N_{,i}}{N}$ from
(\ref{ro}), (\ref{pe}) into (\ref{ls1}), (\ref{ls2}) respectively,
we obtain one equation for $\dot{\rho}$ and one for $\rho_{,i}$.
Then, for the case (i) with $E^{i}_{j}=\delta^{i}_{j}$, it can be
easily seen that $\rho_{,i}=0$ for any equation of state. This
means that in this case, if one wishes to describe inhomogeneous
perfect fluids, has to include non-local bulk effects.
Inhomogeneity in the observable universe is one way to describe
temperature anisotropies in the CMB spectrum (e.g. through the
Sachs-Wolfe effect \cite{sachs}). In the second case (ii) with
$E^{i}_{j}=diag(+1,+1,-1)$, the above equations for $\dot{\rho}$,
$\rho_{,i}$ do not necessarily imply vanishing $\rho_{,i}$\,. If
these equations are solved for $\rho(t,x^{i})$ under some
conditions for the geometry and for some equation of state, then
equations (\ref{ro}), (\ref{pe}) and the other dynamical Einstein
equations will presumably supply the whole cosmological solution.
It is probable that the equation of state should not be given a
priori, but the compatibility of the system would supply some
rather complicated equation of state (as e.g. the solution given
in \cite{bergh}). Thus, even we have not proven the existence of
brane inhomogeneous solutions in an $(A)dS_{5}$ or Minkowski bulk,
we have shown that the conservation equations of the common matter
does not exclude such a possibility. This is in contrast to the
usual brane cosmologies (without including $^{(4)}R$), where the
conservation of energy on the brane in an $(A)dS_{5}$ or Minkowski
bulk enforces the existence of only spatially homogeneous
universes \cite{maeda}.
\par
One can, at this point, look at the right hand side of equations
(\ref{11}), (\ref{12}) of next section, where $V$ and $G$ are
given by (\ref{potential1})-(\ref{force2}). These are the matter
terms appeared for a perfect fluid of case (i) (equations
(\ref{4l1})) in the full general relativity cosmological case (not
only in the spatially homogeneous spacetimes of next section). In
the case where $^{(4)}R$ is not included, in the past history of a
universe, i.e. when $\kappa_{4}^{2}\rho \gg \alpha^{2},
\Lambda_{4}, \Lambda_{5}$, the dominant matter terms are
$\kappa_{4}^{2} \rho$, $\kappa_{4}^{2} p$. Including $^{(4)}R$,
the additional term $\alpha \kappa_{4}\sqrt{\rho}$ emerges
(roughly speaking this is correct when the equation of state has
$p/\rho=w=$constant). But, in the above era of matter domination,
this last term is too small relative to the previous ones and can
be ignored. Thus, we are remained with common general relativity
without any modified matter terms. To estimate the interval of
validity of this approximation we integrate equation (\ref{ro})
for $p=w\rho$ ($-1 < w \leq 1$), i.e.
$\rho=\rho_{1}(x^{i})\,\gamma\,^{-\frac{w+1}{2}}$ ($\rho_{1}$
integration function), and then it arises that one must have
$\gamma \ll
(\frac{\kappa_{4}^{2}\rho_{1}}{\alpha^{2}})^{\frac{2}{w+1}}$. This
value supplies an upper bound estimate of the volume scale factor
of the universe for the usual dynamics to be valid. Obviously, if
$M_{5}$ is small enough, we can revive standard dynamics up to any
desired cosmological scale. The same arguments are also seen to be
true for case (ii).

\section*{3 \,\,\,An anisotropic example}
\hspace{0.8cm} Since we have found the Einstein equations
(\ref{einstein}) governing the modified dynamics of the brane, we
can attempt to find new classes of brane solutions. Due to the
metric (\ref{lineelement}), the system (\ref{einstein}) is
equivalent to the following set of equations (in the rotated
basis supposed that it is a coordinate one):
\be
k_{j\,|\,i}^{i}-(k_{i}^{i})_{\,|\,j}=\varepsilon_{t}\,\kappa_{n}^{2}\,N\,\,^{(n)}T_{j}^{0}\,\,\,\,\,\,\,\,\,,\,\,\,\,\,\,\,\,\,k_{j}^{i}\equiv\frac{1}{2N}\,g^{il}\dot{g}_{lj}\,,
\label{equ0}
\ee
 \be
\dot{g}^{ij}\dot{g}_{ij}+\left(\frac{\dot{\gamma}}{\gamma}\right)^{2}
-4\varepsilon_{t}N^{2}\,\,^{(n-1)}R&=&8\varepsilon_{t}N^{2}\kappa_{n}^{2}\,\,^{(n)}T_{0}^{0}-
8\varepsilon_{t}N^{2}\left(\Lambda_{n}+\varepsilon_{y}\frac{n-1}{n-2}\alpha^{2}\right)+\nn\\&&+\,
\frac{8\varepsilon_{t}\varepsilon_{y}\alpha}{n-2}N^{2}\,\left(\,(n-1)L_{0}^{0}+L_{i}^{i}\,\right),
\label{equ1} \ee \be
&&\ddot{g}_{ij}+\left(\frac{\dot{\gamma}}{2\gamma}-\frac{\dot{N}}{N}\right)\dot{g}_{ij}
-g^{kl}\dot{g}_{ki}\dot{g}_{lj}-2\varepsilon_{t}N^{2}\,\,^{(n-1)}R_{ij}=
\nn\\&&=-2\varepsilon_{t}NN_{|ij}-2\varepsilon_{t}N^{2}\kappa_{n}^{2}\,\left(^{(n)}T_{ij}-
\frac{^{(n)}T}{n-2}g_{ij}\,\right)-2\varepsilon_{t}\varepsilon_{y}\alpha
N^{2}L_{i}^{l}\,g_{lj}+\nn\\&&+\,\frac{2\varepsilon_{t}\varepsilon_{y}
n \alpha}{(n-2)^{2}}N^{2}\,(L_{0}^{0}+L_{l}^{l})\,g_{ij}
-\frac{4\varepsilon_{t}}{n-2}N^{2}\,\left(\Lambda_{n}+\varepsilon_{y}\frac{n-1}{n-2}\alpha^{2}\,\right)\,g_{ij}\,,
\label{equ2} \ee where $|$ stands for the covariant
differentiation with respect to $g_{ij}$.
\par
Hereafter, we focus on cosmological situations. If the
$(n-1)$-dimensional spacelike surfaces foliating the brane are
homogeneous spaces, i.e. there exists a $(n-1)$-dimensional
isometry group of motions acting on each such surface, then we
have a spatially homogeneous brane. More precisely, the group is
assumed to be simply connected and the spacelike surfaces can be
identified with the group by singling out a point on these as the
identity of the group. In this case, there are $n-1$ basis
one-forms $\sigma^{\hat{\alpha}}(x^{i})$,
$\hat{\alpha}=1,...,n-1$, such that
$d\sigma^{\hat{\alpha}}=-C^{\hat{\alpha}}_{\hat{\beta}\hat{\gamma}}\sigma^{\hat{\beta}}\wedge\sigma^{\hat{\gamma}}$
with $C^{\hat{\alpha}}_{\hat{\beta}\hat{\gamma}}$ being the
structure constants of the corresponding isometry group
\cite{ryan}. Furthermore, a general homogeneous tensor field (i.e.
invariant under the motion of the isometry group)
$\Omega^{i}_{j}(t,x)$ (e.g. $g_{ij}$, $k_{j}^{i}$) can be
decomposed as
$\Omega^{\hat{\alpha}}_{\hat{\beta}}(t)\,\sigma^{i}_{\hat{\alpha}}(x)\,\sigma^{\hat{\beta}}_{j}(x)$,
where $\sigma^{i}_{\hat{\alpha}}$ is the inverse matrix of
$\sigma^{\hat{\alpha}}_{i}$. In what follows we focus on the
situation with $\mathcal{T}^{i}_{j}$ being isotropic and also the
case (i) of $n=4$ holding. Then, the previous equations
(\ref{equ0}), (\ref{equ1}), (\ref{equ2}) (for the most interesting
case $\varepsilon_{y}=-\varepsilon_{t}=1$) get the form: \be
k_{\hat{\alpha}}^{\hat{\mu}}C_{\hat{\mu}\hat{\nu}}^{\hat{\nu}}-
k_{\hat{\nu}}^{\hat{\mu}}C_{\hat{\alpha}\hat{\mu}}^{\hat{\nu}}=-\frac{1}{2}\,\kappa_{4}^{2}\,N\,\,^{(4)}T_{\hat{\alpha}}^{0}\,,
\label{eq0} \ee \be
\dot{\gamma}^{\hat{\alpha}\hat{\beta}}\dot{\gamma}_{\hat{\alpha}\hat{\beta}}
+\left(\frac{\dot{\tilde{\gamma}}}{\tilde{\gamma}}\right)^{2}
+4N^{2}\,\,^{(3)}R&=&-8N^{2}\kappa_{4}^{2}\,\,^{(4)}T_{0}^{0}+
8N^{2}\,\left(\Lambda_{4}+\frac{3}{2}\alpha^{2}\right)-\nn\\&&-12\alpha
N^{2}\,(L_{0}^{0}+S)\,,
 \label{eq1}
 \ee
 \be
&&\ddot{\gamma}_{\hat{\alpha}\hat{\beta}}+\left(\frac{\dot{\tilde{\gamma}}}{2\tilde{\gamma}}-\frac{\dot{N}}{N}\right)\dot{\gamma}_{\hat{\alpha}\hat{\beta}}
-\gamma^{\hat{\mu}\hat{\nu}}\dot{\gamma}_{\hat{\mu}\hat{\alpha}}\dot{\gamma}_{\hat{\nu}\hat{\beta}}+2N^{2}\,\,^{(3)}R_{\hat{\alpha}\hat{\beta}}=
\nn\\&&=2N^{2}\kappa_{4}^{2}\,\left(^{(4)}T_{\hat{\alpha}
\hat{\beta}}-
\frac{^{(4)}T}{2}\gamma_{\hat{\alpha}\hat{\beta}}\,\right)
+2N^{2}\,\left(\Lambda_{4}+\frac{3}{2}\alpha^{2}\,\right)\,\gamma_{\hat{\alpha}\hat{\beta}}
-\nn\\&&-2\alpha
N^{2}\,(L_{0}^{0}+2S)\,\gamma_{\hat{\alpha}\hat{\beta}} \,,
 \label{eq2}
 \ee
where
$g_{ij}(t,x)=\gamma_{\hat{\alpha}\hat{\beta}}(t)\,\sigma^{\hat{\alpha}}_{i}(x)\,\sigma^{\hat{\beta}}_{j}(x)$,
$\tilde{\gamma}=det(\gamma_{\hat{\alpha}\hat{\beta}})$ and
$L_{0}^{0}$, $S$ are given by (\ref{4l1}).
\par
For a spatially homogeneous brane, the unique conservation
equation (\ref{ro}), for a perfect fluid matter content, is
integrated to \be
\rho(t)=\rho_{1}\,\tilde{\gamma}\,^{-\frac{w+1}{2}}\,,
\label{consol} \ee where $\rho_{1}$ is a constant of integration.
When, additionally, $\overline{\textsf{E}}^{\,\mu}_{\,\nu}=0$, the
previous equations (\ref{eq1}), (\ref{eq2}) are equivalent to the
following system: \be
\dot{\gamma}^{\hat{\alpha}\hat{\beta}}\dot{\gamma}_{\hat{\alpha}\hat{\beta}}
+\left(\frac{\dot{\tilde{\gamma}}}{\tilde{\gamma}}\right)^{2}
+4N^{2}\,\,^{(3)}R=4N^{2}\,(\,2\Lambda_{4}+3\alpha^{2}+2\,\kappa_{4}^{2}\,\rho+V(\tilde{\gamma})\,)\,,
\label{11} \ee \be
\ddot{\gamma}_{\hat{\alpha}\hat{\beta}}+\left(\frac{\dot{\tilde{\gamma}}}{2\tilde{\gamma}}-\frac{\dot{N}}{N}\right)\dot{\gamma}_{\hat{\alpha}\hat{\beta}}
&-&\gamma^{\hat{\mu}\hat{\nu}}\dot{\gamma}_{\hat{\mu}\hat{\alpha}}\dot{\gamma}_{\hat{\nu}\hat{\beta}}+2N^{2}\,\,^{(3)}R_{\hat{\alpha}\hat{\beta}}=\nn\\&&=
N^{2}\,(\,2\Lambda_{4}+3\alpha^{2}+\kappa_{4}^{2}\,(\rho-p)+G(\tilde{\gamma})\,)\,
\gamma_{\hat{\alpha}\hat{\beta}}\,,
 \label{12}
 \ee
 where
 \be
V(\tilde{\gamma})=\mp
\sqrt{3}\,\alpha\,\sqrt{4\Lambda_{4}-2\Lambda_{5}+3\alpha^{2}+4\kappa_{4}^{2}\,\rho}\,\,,
\label{potential1} \ee \be G(\tilde{\gamma})=\mp
\sqrt{3}\,\alpha\,\frac{4\Lambda_{4}-2\Lambda_{5}+3\alpha^{2}+\kappa_{4}^{2}(3\rho-p)}
{\sqrt{4\Lambda_{4}-2\Lambda_{5}+3\alpha^{2}+4\kappa_{4}^{2}\,\rho}}\,\,,
\label{force1} \ee
  the above $-$ sign (resp. $+$) holding for the $+$ sign of
 equations (\ref{4l1}) and $\mathcal{T}_{0}^{0}+3\tau+3\alpha^{2}>
 0$ (resp. $-$ of (\ref{4l1}) and $\mathcal{T}_{0}^{0}+3\tau+3\alpha^{2}<
 0$),
\newline
  or
 \be
V(\tilde{\gamma})=\mp
\frac{3\sqrt{3}\,\alpha\,\kappa_{4}^{2}\,(\rho+p)}{\sqrt{4\Lambda_{4}-2\Lambda_{5}+3\alpha^{2}+4\kappa_{4}^{2}\,\rho}}
\,\,, \label{potential2} \ee \be G(\tilde{\gamma})=\mp\,
\alpha\,\frac{-4\Lambda_{4}+2\Lambda_{5}-3\alpha^{2}+\kappa_{4}^{2}(5\rho+9p)}
{\sqrt{3}\,\sqrt{4\Lambda_{4}-2\Lambda_{5}+3\alpha^{2}+4\kappa_{4}^{2}\,\rho}}
 \,\,,
 \label{force2}
 \ee
where now, the
 $-$ sign (resp. $+$) of (\ref{potential2}), (\ref{force2}) holds for the $+$ sign of
 equations (\ref{4l1}) and $\mathcal{T}_{0}^{0}+3\tau+3\alpha^{2}<
 0$ (resp. $-$ of (\ref{4l1}) and $\mathcal{T}_{0}^{0}+3\tau+3\alpha^{2}>
 0$). Finally, equation (\ref{eq0}) is written as
 \be
\gamma^{\hat{\beta}\hat{\delta}}(C_{\hat{\alpha}\hat{\beta}}^{\hat{\rho}}\dot{\gamma}_{\hat{\rho}\hat{\delta}}-
C_{\hat{\beta}\hat{\rho}}^{\hat{\rho}}\dot{\gamma}_{\hat{\alpha}\hat{\delta}})=0\,.
\label{linea}
\ee
 \par
 The previous systems of equations can be solved as in
 conventional four-dimensional general relativity, by choosing
 some specific isometry group and adopting an ansatz for the
 three-metric $\gamma_{\hat{\alpha}\hat{\beta}}$. We will proceed
 giving an example,
 namely the anisotropic generalization of the open FRW universe,
 known as Bianchi type V geometry. This is characterized by the
 structure constants $C_{\hat{1}\hat{3}}^{\hat{1}}=C_{\hat{2}\hat{3}}^{\hat{2}}=1/2$, other
 combinations vanish. Choose the temporal gauge $N=\sqrt{\tilde{\gamma}}$.
 Then, equations (\ref{12}) become
 \be
\ddot{\gamma}_{\hat{\alpha}\hat{\beta}}-\gamma^{\hat{\mu}\hat{\nu}}\dot{\gamma}_{\hat{\mu}\hat{\alpha}}\dot{\gamma}_{\hat{\nu}\hat{\beta}}
+\bar{G}(\tilde{\gamma})\gamma_{\hat{\alpha}\hat{\beta}}=0\,,
 \label{finaldynamical}
 \ee
where \be
\bar{G}(\tilde{\gamma})=-4\tilde{\gamma}^{\frac{2}{3}}-\tilde{\gamma}\,(2\Lambda_{4}+3\alpha^{2}+\kappa_{4}^{2}(\rho-p)+G(\tilde{\gamma}))\,.
\label{gbar} \ee The trace of equations (\ref{finaldynamical})
gives \be
\left(\frac{\dot{\tilde{\gamma}}}{\tilde{\gamma}}\right)^{.}+3\bar{G}(\tilde{\gamma})=0\,,
\label{trace} \ee which has a first integral \be
\textit{s}\equiv\left(\frac{\dot{\tilde{\gamma}}}{\tilde{\gamma}}\right)^{2}+U(\tilde{\gamma})=constant\,,
\label{integral} \ee where \be
U(\tilde{\gamma})=6\int\frac{\bar{G}(\tilde{\gamma})}{\tilde{\gamma}}d\tilde{\gamma}\,.
\label{u} \ee Making the conformal transformation \cite{kofinas}
\be
\varpi_{\hat{\alpha}\hat{\beta}}\equiv\tilde{\gamma}^{-1/3}\,\gamma_{\hat{\alpha}\hat{\beta}}\,,
\label{conformal} \ee equations (\ref{finaldynamical}), due to
(\ref{trace}), become \be
\ddot{\varpi}_{\hat{\alpha}\hat{\beta}}-\varpi^{\hat{\mu}\hat{\nu}}\dot{\varpi}_{\hat{\mu}\hat{\alpha}}\dot{\varpi}_{\hat{\nu}\hat{\beta}}=0\,,
\label{zeropot} \ee which are integrated to \be
\dot{\varpi}_{\hat{\alpha}\hat{\beta}}=\vartheta_{\hat{\alpha}}^{\hat{\delta}}\,
\varpi_{\hat{\delta}\hat{\beta}}\,\,, \label{integrals} \ee with
$\vartheta_{\hat{\alpha}}^{\hat{\beta}}$ integration constants.
Since det$(\varpi_{\hat{\alpha}\hat{\beta}})=1$, it is
$\vartheta_{\hat{\alpha}}^{\hat{\alpha}}=0$.
\par
We consider a diagonal metric
$\gamma_{\hat{\alpha}\hat{\beta}}=diag(\gamma_{\hat{1}\hat{1}},\gamma_{\hat{2}\hat{2}},\gamma_{\hat{3}\hat{3}})$
and then, we have, due to the linear constraints (\ref{linea}) that
$\tilde{\gamma}=\gamma_{\hat{3}\hat{3}}^{3}\Leftrightarrow
\gamma_{\hat{1}\hat{1}}\gamma_{\hat{2}\hat{2}}=\gamma_{\hat{3}\hat{3}}^{2}$.
System (\ref{integrals}) supplies
$\vartheta_{\hat{1}}^{\hat{1}}\vartheta_{\hat{2}}^{\hat{2}}\neq0$, while
all other constants $\vartheta_{\hat{\alpha}}^{\hat{\beta}}$
vanish. Thus,
$\vartheta_{\hat{2}}^{\hat{2}}=-\vartheta_{\hat{1}}^{\hat{1}}$.
Then, system (\ref{integrals}), combined with the quadratic equation
(\ref{11}) and the first integral (\ref{integral}), supply, when
(\ref{potential1}) and (\ref{force1}) hold, the following
algebraic relation between the constants of integration:
\be
(\vartheta_{\hat{1}}^{\hat{1}})^{2}=\frac{1}{3}\textit{s}\,.
\label{constrain}
\ee
The other set of equations
(\ref{potential2}), (\ref{force2}) does not give compatibility.
Equation (\ref{integral}), when (\ref{potential1}), (\ref{force1})
hold, gives the following explicit expression for the mean
expansion rate of the described universe
 \be
\left(\frac{\dot{\tilde{\gamma}}}{\tilde{\gamma}}\right)^{2}=\textit{s}+36\,\tilde{\gamma}^{\frac{2}{3}}+
12\kappa_{4}^{2}\,\rho_{1}\,\tilde{\gamma}^{\frac{1-w}{2}}&+&6\tilde{\gamma}\,
\Big(\,2\Lambda_{4}+3\alpha^{2}\mp\nn\\&&\sqrt{3}\,\alpha\sqrt{4\Lambda_{4}-2\Lambda_{5}+3\alpha^{2}+4\kappa_{4}^{2}\,\rho_{1}\,\tilde{\gamma}^{-\frac{1+w}{2}}}\,\Big)\,.
\label{explicit} \ee To conclude, after having found
$\tilde{\gamma}(t)$ from (\ref{explicit}), the solution of our
system is \be
\gamma_{\hat{1}\hat{1}}(t)=e^{\sqrt{\frac{s}{3}}\,t}\,\,\tilde{\gamma}^{\frac{1}{3}}(t)
\label{finsola} \ee \be
\gamma_{\hat{2}\hat{2}}(t)=e^{-\sqrt{\frac{s}{3}}\,t}\,\,\tilde{\gamma}^{\frac{1}{3}}(t)\,,
\label{finsolb} \ee This is the unique perfect fluid solution for
$n=4$ and $E_{j}^{i}=\delta_{j}^{i}$. This solution contains 2
(positive) essential constants $s,\rho_{1}$ (plus three parameters
$w,\Lambda_{4},\Lambda_{5}$) and corresponds to the solution of
4-dimensional general relativity found in \cite{ellis, ruban} (see
also \cite{wainwright}). The difference lies in equation
(\ref{explicit}) and basically in the presence of the terms
multiplied by $\alpha$. For $s=\Lambda_{4}=\Lambda_{5}=0$, the
above solution reduces to the isotropic solution found in
\cite{deffayet}. As it is seen from (\ref{explicit}),
(\ref{finsola}), (\ref{finsolb}) the solution found is always
singular, while for some range of its parameters it is bounded
from above. When $\tilde{\gamma}$ is extendible to infinity,
equation (\ref{explicit}) for $\tilde{\gamma}$ becomes in this
interval the common de Sitter equation, i.e. the mean scale factor
grows exponentially with proper time with an effective
cosmological constant $\Lambda_{4}+\frac{3}{2}\alpha^{2}\mp
\frac{\sqrt{3}}{2}\alpha
\sqrt{4\Lambda_{4}-2\Lambda_{5}+3\alpha^{2}}$\,. Even for
$\Lambda_{4}=\Lambda_{5}=0$, a cosmological constant
$\frac{3}{2}\alpha^{2}$ remains for the $+$ sign solution, a fact
first observed in \cite{shtanov}; for the $-$ sign, the above
effective cosmological constant vanishes and the solution enters a
fully 5D regime. In $AdS_{5}$ bulk with
$|\Lambda_{5}|=\frac{2}{3}(\frac{\Lambda_{4}}{\alpha})^{2}$
($\Lambda_{4} \neq 0$) the effective cosmological constant
vanishes (for suitable $\pm$ sign) and the solution enters once
more a 4D regime with $\kappa_{4}^{2}$ replaced by
$\kappa_{4}^{2}\,(1+\frac{3\alpha^{2}}{2\Lambda_{4}})^{-1}$. In
this $AdS_{5}$ bulk we can find from equation (\ref{trace}), using
also (\ref{force1}), (\ref{gbar}) and (\ref{explicit}), the
following second order differential equation for the mean scale
factor $l \equiv \tilde{\gamma}^{\frac{1}{6}}$ of the universe
with respect to the proper time $t_{P}$\,: \be
2l^{5}\frac{d^{2}l}{dt_{P}^{2}}=-\frac{s}{9}+\alpha^{2}\,l^{3(1-w)}\,
\Big(1+\frac{2\Lambda_{4}}{3\alpha^{2}}\Big)\,\,\Big[&-&(1+3w)\,\beta\,\Big(1+\frac{2
\Lambda_{4}}{3\alpha^{2}}\Big)
\,+\,l^{3(1+w)}-\nn\\&&-\,\frac{l^{3(1+w)}+(1-3w)\beta}{\sqrt{1+4\,\beta\,l^{-3(1+w)}}}\,\,\Big]\,,
\label{secondder} \ee where $\beta \equiv
\frac{\kappa_{4}^{2}\rho_{1}}{3\alpha^{2}}(1+\frac{2\Lambda_{4}}{3\alpha^{2}})^{-2}>0$.
The parameters $\beta, \Lambda_{4}$ are independent of each other
since $\rho_{1}, \Lambda_{4}$ are also independent. By choosing
$\beta$ (for an equation of state with $w>-\frac{1}{3}$), we can
make the difference of the last two terms in the above bracket
positive by simply taking
$l>(\frac{(1-3w)^{2}\beta}{2(1+3w)})^{\frac{1}{3(1+w)}}$. For such
an $l$, say $l_{\ast}$, we can certainly find a $\Lambda_{4}$
(with $2\Lambda_{4}+3\alpha^{2}>0$) such that the whole bracket
being positive; further, there exists an $s$ (e.g. $s=0$) such
that $\frac{d^{2}l}{dt_{P}^{2}}|_{l_{\ast}}>0$. Thus, we have
found a brane solution which emerges as a four-dimensional general
relativistic solution, possesses in the future an accelerating
phase, and finally reduces to a 4D perfect-fluid solution with
modified Newton's constant. Since $w>-\frac{1}{3}$, this solution
does not possess event horizons. This is important in relation to
the problems encountered by string theory to define a set of
observable quantities analogous to an $S$-matrix in ordinary de
Sitter spaces (see e.g. \cite{banks}).
\par
Anisotropic solutions such as the above one, since they contain
$M_{5}$ as a parameter, may serve for relating the five
dimensional Planck mass to the shear/expansion parameter, using the
quadrupole moment $\alpha_{2}=10^{-5}$ of the temperature pattern of
the CMB radiation.

\section*{4 \,\,\,Conclusions}
\hspace{0.8cm} We have studied the dynamics of the most general
3-brane when the intrinsic curvature term is added in the bulk
action. This term is known to cause dramatic changes on the
propagators studied within a fixed - not cosmological -
background. We have reduced the modified dynamics of the brane to
a usual Einstein dynamics coupled to a well-defined modified
matter content. In particular, the total energy-momentum tensor of
these equations is split into the common four-dimensional
energy-momentum tensor plus additional terms, which are all
multiplied by one of the characteristic scales of the theory, i.e.
$1/r_{c}=M_{5}^{3}/M_{4}^{2}$. These additional terms are
basically square roots of the various matter terms, while
non-local bulk effects onto the brane, carried by the electric
part of the higher dimensional Weyl tensor, are also included.
Brane dynamics is made closed by setting boundary conditions (for
the propagation equations in the bulk space) to the non-local
terms on the brane. From a cosmological viewpoint, the various
non-conventional matter terms can be dropped in the first era of
the universe evolution characterized by some volume scale factor
much smaller than a positive power of $r_{c}$. Inhomogeneous
perfect fluid solutions were seen not to be necessarily
incompatible with exact $(A)dS$ or Minkowski bulks, in contrast to
the case of brane cosmologies not containing the $^{(4)}R$ term.
The above form for the induced equations is useful in order to
find new brane solutions following the methods of general
relativity theory. As an application of this, we have found a new
brane cosmological solution for an anisotropic brane, where the
spacelike surfaces admit the isometry group of Bianchi type V
geometry, i.e. \be
ds_{(4)}^{2}=-\tilde{\gamma}(t)dt^{2}+\tilde{\gamma}(t)^{\frac{1}{3}}
\left[e^{\sqrt{\frac{s}{3}}\,t}\,(\sigma^{\hat{1}})^{2}+
e^{-\sqrt{\frac{s}{3}}\,t}\,(\sigma^{\hat{2}})^{2}
+(\sigma^{\hat{3}})^{2}\right]\,, \label{metricfinal} \ee with
$\tilde{\gamma}(t)$ given by equation (\ref{explicit}). Choosing
$s=0$ an isotropic solution arises. Solution (\ref{metricfinal})
approaches a known general relativistic solution in the beginning
of the evolution, but finally it evolves as an inflationary
anisotropic solution. We have not investigated the embedding of
this braneworld in the bulk space. For a fine-tuned
five-dimensional cosmological constant (negative), this solution
eternally evolves as a usual perfect fluid solution (with
non-quintessence equation of state) with no effective
four-dimensional cosmological constant appearing, except that the
effective Newton's constant equals the Newton's constant divided
by $1+\frac{3\alpha^{2}}{2\Lambda_{4}}$; thus, future horizons do
not appear. Before this phase, there exists an accelerating era
which may be in agreement with the present supernovae data
\cite{riess}.
\par
Though the $^{(4)}R$ term arises from first order quantum
corrections, it does alter the conventional behavior towards the
initial singularity (whenever this encounters). Possible
violations of the energy conditions may occur at later stages of
the evolution. Any inhomogeneous or anisotropic solution found,
based on the modified brane dynamics, would serve to investigate
possible deviations from the de Sitter future evolution, as
suggested by the cosmic no-hair conjecture.
\par
Due to the non-uniqueness of $E_{j}^{i}$, one could assume
that the induced dynamics is not unambiguous. It is better, instead, to
take the point of view that dynamics is well-defined, but the governing equations
are further classified according to which class of solutions we want to pick.
This classification has, of course, to do with the initial conditions of the
system.
\par
The inclusion of next order corrections in a
derivative expansion of the action involves the $^{(5)}R^{2}$ term
in the bulk and the $^{(4)}R^{2}$ term on the brane. One could try
including these terms in an analysis of the induced brane dynamics.
\par
In \cite{corley}, except the $^{(4)}R$ term, two more additional
terms, namely $K_{\mu}^{\mu}K_{\nu}^{\nu}$ and
$K_{\mu}^{\nu}K_{\nu}^{\mu}$ proved to be necessary in the low
energy effective action to reproduce the tree level string
amplitude corresponding to scattering a massless closed string
field off a bosonic $Dp$-brane. It would be interesting to get
modified boundary equations instead of (\ref{eqn1}), when these
two extra terms are present. Equations (\ref{eqn1}) arise
alternatively, including in the action (\ref{action}) the standard
Gibbons-Hawking term \cite{gibbons} and making the variation with
respect to $g_{AB}$ \cite{chamblin, shtanov}. It seems, however,
that the inclusion of these two terms does not supply a well
defined variational problem.

\section*{Acknowledgements}
We wish to thank T. Christodoulakis, A. Daniilidis, A. Kehagias,
E. Kiritsis, E. Papantonopoulos, I. Pappa and N. Tetradis for helpful comments and discussions.



 \end{document}